\documentstyle[12pt]{article}

\begin{document}
{\sf \begin{center} \noindent
{\Large \bf Ricci curvature and geodesic flows stability in Riemannian twisted flux tubes}\\[3mm]

by \\[0.3cm]

{\sl L.C. Garcia de Andrade}\\

\vspace{0.5cm} Departamento de F\'{\i}sica
Te\'orica -- IF -- Universidade do Estado do Rio de Janeiro-UERJ\\[-3mm]
Rua S\~ao Francisco Xavier, 524\\[-3mm]
Cep 20550-003, Maracan\~a, Rio de Janeiro, RJ, Brasil\\[-3mm]
Electronic mail address: garcia@dft.if.uerj.br\\[-3mm]
\vspace{2cm} {\bf Abstract}
\end{center}
\paragraph*{}
Ricci and sectional curvatures of twisted flux tubes in Riemannian
manifold are computed to investigate the stability of the tubes. The
geodesic equations are used to show that in the case of thick tubes,
the curvature of planar (Frenet torsion-free) tubes have the effect
ct of damping the flow speed along the tube. Stability of geodesic
flows in the Riemannian twisted thin tubes (almost filaments),
against constant radial perturbations is investigated by using the
method of negative sectional curvature for unstable flows. No
special form of the flow like Beltrami flows is admitted, and the
proof is general for the case of thin tubes. It is found that for
positive perturbations and angular speed of the flow, instability is
achieved , since the sectional Ricci curvature of the twisted tube
metric is negative.\vspace{0.5cm} \noindent {\bf PACS numbers:}
\hfill\parbox[t]{13.5cm}{02.40.Hw:Riemannian geometries}

\newpage
\newpage
 \section{Introduction}
 The stability of geodesic flows have been recently investigated by Kambe \cite{1} by making use of the technique
 of Ricci sectional curvature \cite{2}, where the negative sectional
 curvature indicates instability of the flow, while positivity or
 nul indicates stability. In the case of instability the geodesics
 deviate from the perturbation of the fluid. Following the work of D. Anosov \cite{3} on the perturbation in geodesic flows in
 three-dimensional Riemannian geometry, in this paper the sectional Riemann curvature of the geodesic flow
 for a Riemannian flux tube \cite{4,5}, where the axis of the tube flow possesses Frenet curvature and
 torsion. In the approximation of a thin tube where the radius of the tube is almost null, we show that the flows are unstable, against orthogonal perturbations, which is equivalently
 due to the negativity of the sectional Ricci sectional curvature. Throughout the paper the ellegant coordinate-free language of differential geometry \cite{6}
 is used. The paper is organized as follows: Section II presents a
 brief review of Riemannian geometry in the coordinate free
 language.
 Section III presents the geodesic flow computation of the Christoffel symbols for the thick flux tube, where we show that the
 curvature of the tube axihe speed of the flow. Section IV presents the computation of the instability of Riemannian tube
 flow. Section V presents the conclusions.
 \section{Ricci and sectional Riemann curvatures}
 In this section we make a brief review of the differential geometry of surfaces in coordinate-free language.
 The Riemann curvature is defined by
 \begin{equation}
 R(X,Y)Z:={\nabla}_{X}{\nabla}_{Y}Z-{\nabla}_{Y}{\nabla}_{X}Z-{\nabla}_{[X,Y]}Z\label{1}
 \end{equation}
where $X {\epsilon} T\cal{M}$ is the vector representation which is
defined on the tangent space $T\cal{M}$ to the manifold $\cal{M}$.
Here ${\nabla}_{X}Y$ represents the covariant derivative given by
\begin{equation}
{\nabla}_{X}{Y}= (X.{\nabla})Y\label{2}
 \end{equation}
which for the physicists is intuitive, since we are saying that we
are performing derivative along the X direction. The expression
$[X,Y]$ represents the commutator, which on a vector basis frame
${\vec{e}}_{l}$ in this tangent sub-manifold defined by
\begin{equation}
X= X_{k}{\vec{e}}_{k}\label{3}
\end{equation}
or in the dual basis ${{\partial}_{k}}$
\begin{equation}
X= X^{k}{\partial}_{k}\label{4}
\end{equation}
can be expressed as
\begin{equation}
[X,Y]= (X,Y)^{k}{\partial}_{k}\label{5}
\end{equation}
In this same coordinate basis now we are able to write the curvature
expression (\ref{1}) as
\begin{equation}
R(X,Y)Z:=[{R^{l}}_{jkp}Z^{j}X^{k}Y^{p}]{\partial}_{l}\label{6}
\end{equation}
where the Einstein summation convention of tensor calculus is used.
The expression $R(X,Y)Y$ which we shall compute bellow is called
Ricci curvature. The sectional curvature which is very useful in
future computations is defined by
\begin{equation}
K(X,Y):=\frac{<R(X,Y)Y,X>}{S(X,Y)}\label{7}
\end{equation}
where $S(X,Y)$ is defined by
\begin{equation}
{S(X,Y)}:= ||X||^{2}||Y||^{2}-<X,Y>^{2}\label{8}
\end{equation}
where the symbol $<,>$ implies internal product.
\newpage
\section{Geodesic equations in Riemannian tube metric} In this section we shall consider the twisted flux tube Riemann
metric. The metric $g(X,Y)$ line element can be defined as
\cite{4,5}
\begin{equation}
ds^{2}=dr^{2}+r^{2}d{{\theta}_{R}}^{2}+{K^{2}}(s)ds^{2} \label{9}
\end{equation}
This line element was used previously by Ricca \cite{4} and the
author \cite{5} as a magnetic flux tubes with applications in solar
and plasma astrophysics. This is a Riemannian line element
\begin{equation}
ds^{2}=g_{ij}dx^{i}dx^{j} \label{10}
\end{equation}
if the tube coordinates are $(r,{\theta}_{R},s)$ \cite{4} where
${\theta}(s)={\theta}_{R}-\int{{\tau}ds}$ where $\tau$ is the Frenet
torsion of the tube axis and $K(s)$ is given by
\begin{equation}
{K^{2}}(s)=[1-r{\kappa}(s)cos{\theta}(s)]^{2} \label{11}
\end{equation}
Let us now compute the geodesic equations
\begin{equation}
\frac{dv^{i}}{dt}+{{\Gamma}^{i}}_{jk}v^{j}v^{k}=0 \label{12}
\end{equation}
where $v^{s}=\frac{ds}{dt}$ and $v^{\theta}=\frac{d{\theta}}{dt}$
the Riemann-Christoffel symbols are given by
\begin{equation}
{{\Gamma}^{i}}_{jk}=\frac{1}{2}g^{il}[g_{lj,k}+g_{lk,j}-g_{jk,l}]
\label{13}
\end{equation}
The only nonvanishing components of the Christoffel symbols or
Levi-Civita \cite{6} connections of the flux tube are
\begin{equation}
{{\Gamma}^{2}}_{21}=\frac{1}{r} \label{14}
\end{equation}
\begin{equation}
{{\Gamma}^{2}}_{33}=\frac{-K(s){\kappa}sin{\theta}}{r} \label{15}
\end{equation}
\begin{equation}
{{\Gamma}^{3}}_{11}=\frac{r{\kappa}}{2K}[{\tau}sin{\theta}+\frac{{\kappa}'}{\kappa}cos{\theta}]
\label{16}
\end{equation}
\begin{equation}
{{\Gamma}^{3}}_{33}=K^{-1}{K}'=-{{\Gamma}^{3}}_{11} \label{17}
\end{equation}
A simple example can be given by writing the geodesic equation for
the untwisted tube, where $v_{\theta}=0$, as
\begin{equation}
\ddot{s}+{{\Gamma}^{3}}_{11}[{\dot{r}}^{2}-{\dot{s}}^{2}]=0
\label{18}
\end{equation}
where substitution of the component ${{\Gamma}^{3}}_{11}$ above
yields
\begin{equation}
{d{lnv_{s}}}=-d(ln{\kappa}) \label{19}
\end{equation}
where we use the chain rule of differential calculus,
$\frac{ds}{{v_{s}}^{2}}\frac{dv_{s}}{dt}=\frac{dv_{s}}{v_{s}}$.
Solution of equation (\ref{19}) is
\begin{equation}
v_{s}=v_{0}{\kappa}^{-1} \label{20}
\end{equation}
where $v_{0}$ is an integration constant. This solution tells us
that when the curvature tends to $\infty$ the velocity along the
tube axis vanishes. Physically this means that the curvature acts as
a damping to the flow, along the tube. In the next section we
investigate in some detail the stability of an incompressible or
volume preserving flow, using the method of the sign of the Ricci
sectional curvature.
\section{Ricci sectional curvature and flow stability}
One of the most important features of the investigation of the
stability of flows in the Euclidean manifold ${\cal{E}}^{3}$, is the
comprehension of the fact that the covariant derivative in the flow
curved manifold is given by the gradient operator in curvilinear
coordinates. Thus to compute the Ricci sectional curvature above, we
need to make use of the grad operator in the twisted flux tube
Riemannian metric given by
\begin{equation}
\nabla=[{\partial}_{r},r^{-1}{\partial}_{{\theta}_{R}},K^{-1}{\partial}_{s}]
\label{21}
\end{equation}
Since the axis of the tube undergoes torsion and curvature, we need
some dynamical relations from vector analysis and differential
geometry of curves \cite{7} such as the Frenet frame
$(\vec{t},\vec{n},\vec{b})$ equations
\begin{equation}
\vec{t}'=\kappa\vec{n} \label{22}
\end{equation}
\begin{equation}
\vec{n}'=-\kappa\vec{t}+ {\tau}\vec{b} \label{23}
\end{equation}
\begin{equation}
\vec{b}'=-{\tau}\vec{n} \label{24}
\end{equation}
and the other frame vectors are
\begin{equation}
\vec{e_{r}}=\vec{n}cos{\theta}+\vec{b}sin{\theta} \label{25}
\end{equation}
\begin{equation}
\vec{e_{\theta}}=-\vec{n}sin{\theta}+\vec{b}cos{\theta} \label{26}
\end{equation}
\begin{equation}
{{\partial}_{\theta}}\vec{e_{\theta}}=-\vec{n}[(1+{\tau}^{-1}\kappa){sin{\theta}}+cos{\theta}]-\vec{b}[cos{\theta}+sin{\theta}]
\label{27}
\end{equation}
Let the constant perturbation be given by
\begin{equation}
X={u_{r}}^{1}\vec{e}_{r}\label{28}
\end{equation}
The upper index one in this expression refers to the fact that the
background original value of $u_{r}$ was considered as
${u_{r}}^{0}=0$ to form the tube. The other variable Y is given by
\begin{equation}
Y=u_{\theta}{\vec{e}}_{\theta}+u_{s}\vec{t}\label{29}
\end{equation}
Therefore to compute the Ricci tensor step by step we start by the
term
\begin{equation}{\nabla}_{X}Y=
{u_{r}}^{(1)}{\partial}_{r}[u_{\theta}\vec{e}_{\theta}+u_{s}\vec{t}]
\label{30}
\end{equation}
which vanishes since we addopt here the approximation
${u_{r}}^{(1)}{\partial}_{r}[u_{\theta}]\approx{0}$ along with the
same relation to the radial partial derivative of $u_{r}$. So
\begin{equation}{\nabla}_{Y}{\nabla}_{X}Y\approx{0}
\label{31}
\end{equation}
Now the second term in the Ricci tensor is
\begin{equation}{\nabla}_{X}{\nabla}_{Y}Y=-{u_{r}}^{(1)}r^{-2}u_{\theta}[\frac{(1+\tau)}{\tau}(\vec{n}cos{\theta}+\vec{b}sin{\theta})]
\label{32}
\end{equation}
where we have used the approximation of the thin tube where
$K(s)\approx{1}$ and $r\approx{0}$
\begin{equation} [X.Y]= {u_{r}}^{(1)}[r^{-1}u_{\theta}-{\tau}u_{s}][{\vec{e}}_{r}-{\tau}^{-1}\vec{t}]\label{33}
\end{equation}
which implies that
\begin{equation} {\nabla}_{[X,Y]}Y={u_{r}}^{(1)}[r^{-1}u_{\theta}-{\tau}u_{s}][-{\tau}^{-1}u_{\theta}{\partial}_{s}{\vec{e}}_{\theta}+\kappa\vec{n}]
\label{34}
\end{equation}
The Ricci tensor is
\begin{equation}
R(X,Y)Y=-{u_{r}}^{(1)}r^{-1}u_{\theta}{\tau}^{-1}{\partial}_{s}{\vec{e}}_{\theta}
\label{35}
\end{equation}
\newpage

In the previous computations we have made use of the
imcompressibility of the flow
\begin{equation}
{\nabla}.\vec{u}=0 \label{36}
\end{equation}
which is
\begin{equation}
{\partial}_{s}u_{\theta}={\tau}r{\kappa}{u}_{\theta}\approx{0}
\label{37}
\end{equation}
since $r\approx{0}$ on the RHS of equation (\ref{37}). The sectional
curvature is thus
\begin{equation}
K(X,Y)= \frac{<R(X,Y)Y,X>}{S(X,Y)}=
-\frac{u_{\theta}[1+{\tau}(s){\kappa}cos{\theta}]}{r{u^{(1)}}_{r}[{u_{\theta}}^{2}+{u_{s}}^{2}]}\label{38}
\end{equation}
when the tube is strongly twisted, ${{u}_{\theta}}^{2}>>{u_{s}}^{2}$
thus the sectional Ricci curvature is
\begin{equation}
K(X,Y)= \frac{<R(X,Y)Y,X>}{S(X,Y)}=
-\frac{[1+{\tau}(s){\kappa}cos{\theta}]}{r{u^{(1)}}_{r}u_{\theta}}\label{39}
\end{equation}
when the tube, besides is planar or torsion vanishes the last
expression reduces to
\begin{equation}
K(X,Y)= \frac{<R(X,Y)Y,X>}{S(X,Y)}=
-\frac{1}{r{u^{(1)}}_{r}u_{\theta}}\label{40}
\end{equation}
Note that when both angular velocity and perturbation both keep the
same sign, the sectional curvature $K(X,Y)$ is negative and the flow
along the Riemannian flux tube is unstable. There is singularity in
this sectional Riemannian curvature in $r\approx{0}$.
\section{Conclusions}
An important issue in plasma astrophysics as well as in fluid
mechanics is to know when a fluid, charged or not, is unstable or
not. In this paper we discuss and present incompressible flows and
investigate their stability. Instability is obtained even before the
singularity is achieved.

\end{document}